\documentclass[twocolumn,aps,showpacs,superscriptaddress,floatfix]
{revtex4}
\usepackage[dvips]{graphicx}
\newcommand{\ket}[1]{\left | #1 \right \rangle}
\newcommand{\bra}[1]{\left \langle #1 \right |}

\newcommand{\proj}[1]{\ket{#1} \bra{#1}}
\newcommand{\tr}{\mathrm{Tr}}
\newcommand{\be}{\begin{equation}}
\newcommand{\ee}{\end{equation}}
\begin{document}
\title{%
    Optimal parameter estimation of depolarizing channel}
\author{Masahide Sasaki}
\email{psasaki@crl.go.jp}
\affiliation{Communications Research Laboratory,
    Koganei, Tokyo 184-8795, Japan}
\affiliation{CREST, Japan Science and Technology Agency}
\author{Masashi Ban}
\affiliation{Advanced Research Laboratory, Hitachi Ltd, 1-280, 
    Higashi-Koigakubo, Kokubunnji, Tokyo  
    185-8601, Japan}
\author{Stephen M. Barnett}
\affiliation{Department of Physics and Applied Physics,
    University of Strathclyde, Glasgow G4 0NG, Scotland}
\begin{abstract}
We investigate strategies for estimating a depolarizing 
channel for a finite dimensional system. 
Our analysis addresses the double optimization problem of selecting 
the best input probe state and the measurement strategy that minimizes 
the Bayes cost of a quadratic function. 
In the qubit case, we derive the Bayes optimal strategy for any finite 
number of input probe particles when bipartite entanglement can be 
formed in the probe particles. 
\end{abstract}
%
\date{\today}
\maketitle
%
%
\section{Introduction}
\label{sec:intro}
In order to design a reliable communication system one requires 
a priori knowledge of the property of a channel. 
Precise knowledge of the channel allows us to devise 
appropriate coding, modulation, and filtering schemes. 
In general, the channel property is not stationary, 
so one should first acquire and then track the optimal operating point 
of each device by monitoring the condition of the channel. 
It is important, therefore to know how to estimate the channel 
property in an efficient way, that is, 
as precisely as possible with minimum resources.

A reasonable assumption is that we know that the channel belongs to a 
certain parameterized family, and only the values of the parameters 
are not known. To know them one may input a probe system in an 
appropriate state into the channel and make a measurement on the 
output state.  
Only when an infinite amount of input resource is available, 
one can determine the channel parameters with perfect accuracy. 
In the quantum domain, however, the resource is often restricted for 
various reasons. For example, when one is to monitor a fast quantum 
dymanics at cryogenic temperatures, the input probe power should be 
kept as low as possible so as to  prevent the system from heating 
up while obtaining meaningful data in a short time. 
This restricts the available amount of probe particles. 
Furthermore, preparing the probe in an appropriate quantum state 
is usually an elaborate process. 
Thus to find the efficient estimation strategy relying only on 
a restricted amount of input resource is of practical importance.

In estimating a quantum channel parameter, given a finite amount of 
input resource, 
both the input probe state and the measurement of the output state 
need to be optimized. 
This double maximization problem has been studied in the context of 
estimation of SU($d$) unitary operation~
\cite{Acin01}. 
Estimating a noisy quantum channel has been discussed in the 
literature~
\cite{Fujiwara01,Fischer01,Cirone01}. 
In ref. \cite{Fujiwara01}, the locally unbiased estimator and the 
Cram\'er-Rao bound are extensively discussed for the depolarizing 
channel for a qubit system. The locally optimal strategy, which 
achieves the Cram\'er-Rao bound at a local point of the parameter 
space was derived when two qubits at most are used. 
This result would be useful in the limit of large ensemble of the 
input probe. In such a limit, of course, 
one can establish the channel parameter with a very high degree of 
accuracy. 
To improve the rate at which 
the estimation accuracy grows with the number of probe particles, 
one may first apply some preliminary estimation using a part of probe 
particles to establish the most likely value of the parameter, 
and then use the locally optimal strategy around this value to get 
the final estimate~
\cite{Barndorff98,Gill00,Hayashi02}. 
Refs. \cite{Fischer01,Cirone01} focus on several noisy qubit channels. 
They study some reasonable, although not optimal, strategies based on 
maximum likelyhood estimator, and derive the asymptotic behavior of 
the cost as a function of the number of input probe qubits.

In contrast, we are concerned here with the Bayes optimal strategy 
which minimizes the \textit{average} cost. 
The scenario we have in mind is that one has no particular 
knowledge about the a priori parameter distribution, and 
the available number of probe particles is strictly limited.
We then take into account the possibility of rather large errors. 
We seek the strategy that works equally well for all 
possible values of the parameter on average, 
that is, the strategy which is more universal for various possible 
situations.

It seems difficult for us to study this problem for the most general 
probe state. 
In this paper we deal with the depolarizing channel by assuming that 
we dispose of $M$ pairs of probe particles and 
only bipartite entanglement can be formed in each pair. 
This might be a practically sensible assumption from the view point of 
optical implementation given current technology.  
Our problem is to find the best estimation strategy to 
minimize the average cost. 
We consider the quadratic of a cost function.

\section{Qubit case}
\label{sec:qubit}

Let $\hat\rho$ be a density operator in the 2 dimensional Hilbert 
space ${\cal H}_2$. 
The depolarizing channel ${\cal L}_\theta$ maps a density operator 
$\hat\rho$ to a density operator which is a mixture of $\hat\rho$ 
and the maximally mixed state, 
\be
{\cal L}_\theta \hat\rho = \theta \hat\rho 
                         + \frac{1-\theta}{2}\hat I. 
\ee
The parameter $\theta$ represents the degree of randomization of 
polarization. For the map ${\cal L}_\theta$ to be completely 
positive, the parameter $\theta$ must lie in the interval 
$-{1\over3}\le\theta\le1$.

Let us start with two qubit systems as the input probe. 
For simplicity we only consider a pure state family of the probe 
$\hat\Psi=\proj\Psi$. 
This may be represented in the Schmidt decomposition  
\be
\ket{\Psi}=\sqrt{x}\ket0\otimes\ket{e_0}
          +\sqrt{1-x}\ket1\otimes\ket{e_1}, 
\ee
where $\{\ket0,\ket1\}$ and $\{\ket{e_0},\ket{e_1}\}$ are orthonormal 
basis sets for the first and second probe particle, respectively. 
What is the best way to use this state? 
There are two possibilities to consider; 
\begin{itemize}
\item[(a)] \enskip
Input one qubit of the pair into the channel keeping the other 
untouched leading to the output state 
\be
\hat\Psi_1(\theta)
    \equiv({\cal L}_\theta\otimes\hat I)\proj{\Psi}, 
\ee
\item[(b)] \enskip
Input both qubits into the channel and have the output state 
\be
\hat\Psi_2(\theta)
    \equiv({\cal L}_\theta\otimes{\cal L}_\theta)\proj{\Psi}.  
\ee
\end{itemize}
A measurement is described by a probability operator measure (POM)
$\hat\Pi(\theta)$~
\cite{Helstrom_QDET,Holevo_book}, 
also referred to as a positive operator valued measure (POVM) 
\cite{Peres_book}.
The average cost for the quadratic cost function is given by 
\be
\bar C_i(x)
  =\int_{-{1\over3}}^1d\tilde\theta 
   \int_{-{1\over3}}^1d\theta 
   (\tilde\theta-\theta)^2 z(\theta) 
   \tr\left[ \hat\Pi(\tilde\theta) \hat\Psi_i(\theta) \right],  
\ee
where 
$z(\theta)$ is the a priori probability distribution of $\theta$, 
and $\int_{-{1\over3}}^1d\tilde\theta \hat\Pi(\tilde\theta)=\hat I$. 
It is assumed that we have no a priori knowledge about $\theta$, 
that is, $z(\theta)={3\over4}$. 
Given the channel ${\cal L}_\theta$, we are to find the optimal 
probe $\ket{\Psi}$ and the POM $\hat\Pi(\theta)$ minimizing 
the average cost $\bar C(x)$.

It is convenient to introduce the \textit{risk} operator 
\begin{eqnarray}
\hat W(\theta)
&=&{3\over4}\int_{-{1\over3}}^1d\theta' 
   (\theta-\theta')^2 \hat\Psi_i(\theta'), \\
&=&\hat W^{(2)} - 2\theta \hat W^{(1)} + \theta^2 \hat W^{(0)},
\end{eqnarray}
where 
$\hat W^{(k)}\equiv{3\over4}\int_{-{1\over3}}^1d\theta
                      \theta^k \hat\Psi_i(\theta)$. 
The average cost is then  
\be
\bar C(x)=\tr\hat\Gamma, \quad
\hat\Gamma
\equiv 
\int_{-{1\over3}}^1d\theta \hat\Pi(\theta) \hat W(\theta).
\ee
For a fixed probe state $\ket{\Psi}$, 
the optimal POM $\hat\Pi(\theta)$ is derived 
from the necessary and sufficient conditions to minimize the average 
cost~
\cite{Holevo73_condition,YuenKennedyLax75}: 
\begin{itemize}
\item[(i)] \enskip
$\hat\Gamma=\hat\Gamma^\dagger$, and 
$\left[\hat W(\theta)-\hat\Gamma\right]\hat\Pi(\theta)=0$ 
for all $\theta$, 
\item[(ii)] \enskip
$\hat W(\theta)-\hat\Gamma\ge0$ for all $\theta$. 
\end{itemize}
The optimal solution for a single parameter estimation with a 
quadratic cost is well known~
\cite{Personick71b,Helstrom_QDET}.  
The optimal POM is constructed by finding the eigenstate 
$\ket\theta$ of the \textit{minimizing} operator $\hat\Theta$ 
which is defined by 
\be
\hat\Theta \hat W^{(0)} + \hat W^{(0)} \hat\Theta = 2 \hat W^{(1)},  
\ee
that is, $\hat\Pi(\theta)=\proj\theta$ 
so that $\hat\Theta\ket\theta=\theta\ket\theta$.  
We then have 
$\hat\Gamma=\hat W^{(2)}-\hat\Theta \hat W^{(0)} \hat\Theta$ 
from which the conditions (i) and (ii) are easily verified.

For a discrete system, one can find the optimal POM with 
finite elements. 
Let the spectral decomposition of $\hat W^{(0)}$ for our two-qubit 
system be 
\be\label{W0}
\hat W^{(0)}=\sum_{i=1}^4 \omega_i \proj{\omega_i}. 
\ee
Then the minimizing operator is 
\be
\hat\Theta=\sum_{i,j=1}^4 \frac{2}{\omega_i+\omega_j} 
\proj{\omega_i} \hat W^{(1)} \proj{\omega_j}. 
\label{minmizing_op}
\ee
Let the spectral decomposition of $\hat\Theta$ be 
\be
\hat\Theta=\sum_{i=1}^4 \theta_i \proj{\theta_i}. 
\ee
The optimal POM is then given by  
\be
\hat\Pi(\theta)=\sum_{i=1}^4 \delta(\theta-\theta_i) \proj{\theta_i}. 
\ee
This implies that the measurement has 4 outputs at most and 
we then estimate the channel parameter as one of 4 $\theta_i$'s. 
Before going on to derive the optimal strategies, let us define some 
notations. As seen below the output states $\hat\Psi_i(\theta)$'s 
can be written as a direct sum
\be
\hat\Psi_i(\theta)=\hat\psi_i(\theta)\oplus\hat\phi_i(\theta),  
\ee
where $\hat\psi_i(\theta)$ is in the subspace $\cal{H}_\psi$ spanned 
by 
$\ket{\mu_1}\equiv\ket0\otimes\ket{f_0}$ and 
$\ket{\mu_2}\equiv\ket1\otimes\ket{f_1}$, 
and $\hat\phi_i(\theta)$ in the subspace $\cal{H}_\phi$ spanned by 
$\ket{\nu_1}\equiv\ket0\otimes\ket{f_1}$ and 
$\ket{\nu_2}\equiv\ket1\otimes\ket{f_0}$. 
In the following all 2$\times$2 matrices represent density operators 
in $\cal{H}_\psi$ with 
$\ket{\mu_1}=\left(\begin{array}{c} 1 \\ 0 \end{array}\right)$ and 
$\ket{\mu_2}=\left(\begin{array}{c} 0 \\ 1 \end{array}\right)$.

\noindent
\textbf{Case (a)}:

The output state $\hat\Psi_1(\theta)$ is given by 
\begin{eqnarray}
\hat\psi_1(\theta)
&=&
{1\over2}
\left[
\begin{array}{cc}
(1+\theta)x & 2\theta\sqrt{x(1-x)} \\
2\theta\sqrt{x(1-x)} & (1+\theta)(1-x)
\end{array}
\right], 
\\
\hat\phi_1(\theta)
&=& \frac{1-\theta}{2} 
  \Bigl[ (1-x)\proj{\nu_1} + x \proj{\nu_2} \Bigr].   
\end{eqnarray}
The elements of the risk operator are 
\begin{eqnarray}
\hat W^{(0)}
&=&{1\over3}
   \Bigl(
   \left[
         \begin{array}{cc}
         2x & \sqrt{x(1-x)} \\
         \sqrt{x(1-x)} & 2(1-x)
         \end{array}
   \right]
   \oplus \hat\varphi_1
   \Bigr),  
\\
\hat W^{(1)}&=&{1\over{27}}
   \Bigl(
   \left[
         \begin{array}{cc}
         8x & 7\sqrt{x(1-x)} \\
         7\sqrt{x(1-x)} & 8(1-x)
         \end{array}
   \right]
   \oplus \hat\varphi_1
   \Bigr),  
\\
\hat W^{(2)}&=&{1\over{27}}
   \Bigl(
   \left[
         \begin{array}{cc}
         6x & 5\sqrt{x(1-x)} \\
         5\sqrt{x(1-x)} & 6(1-x)
         \end{array}
   \right]
   \oplus \hat\varphi_1
   \Bigr),  
\end{eqnarray}
where
$\hat\varphi_1=(1-x)\proj{\nu_1} + x \proj{\nu_2}$. 
After a lengthy but straightforward calculation 
(see Appendix \ref{app_a}) 
we have 
\be\label{Theta_a}
\hat\Theta
={2\over9}
   \left[
         \begin{array}{cc}
         1+x & 2\sqrt{x(1-x)} \\
         2\sqrt{x(1-x)} & 2-x
         \end{array}
   \right]
   \oplus {1\over9}\hat I_\phi. 
\ee
To diagonalize $\Theta$ we introduce $r=\sqrt{1+12x(1-x)}$ and 
\be
\mathrm{cos}\gamma=\sqrt{\frac{r-1+2x}{2r}}, 
\quad
\mathrm{sin}\gamma=\sqrt{\frac{r+1-2x}{2r}}. 
\ee
The eigenstates and eigenvalues are then 
\be\label{eigenvec1}
\begin{array}{lll}
\ket{\theta_1}&=\mathrm{cos}\gamma\ket{\mu_1}
               +\mathrm{sin}\gamma\ket{\mu_2}, \quad
              &\theta_1=(3+r)/9, \\
\ket{\theta_2}&=-\mathrm{sin}\gamma\ket{\mu_1}
                +\mathrm{cos}\gamma\ket{\mu_2}, \quad
              &\theta_2=(3-r)/9, \\
\ket{\theta_3}&=\ket{\nu_1}, \quad
              &\theta_3=1/9, \\
\ket{\theta_4}&=\ket{\nu_2}, \quad
              &\theta_4=1/9. 
\end{array} 
\ee
The average cost finally reads  
\be
\bar C_1(x)=\tr(\hat W^{(2)}-\Theta\hat W^{(0)}\Theta)
={8\over{81}}\left[1+(x-{1\over2})^2\right]. 
\ee
This is minimized by the maximally entangled state input 
\be
\ket{\Psi}=\frac{1}{\sqrt2}
\left(\ket0\otimes\ket{f_0}+\ket1\otimes\ket{f_1}\right), 
\ee 
for which $\theta_1={5\over9}$ and 
$\theta_2=\theta_3=\theta_4={1\over9}$. Therefore the optimal 
measurement is actually constructed by the two projectors 
\be\label{optimal POM}
\hat\Pi_1=\proj{\Psi}, \quad \hat\Pi_2=\hat I - \proj{\Psi},  
\ee
with the associated guesses  
$\theta_1={5\over9}$ and $\theta_2={1\over9}$, respectively. 
The minimum average cost is $\bar C_{1\mathrm{min}}={8\over81}$.

\noindent
\textbf{Case (b)}:

The output state 
$\hat\Psi_2(\theta)=\hat\psi_2(\theta)\oplus\hat\phi_2(\theta)$
is given by 
\begin{eqnarray}
\hat\psi_2(\theta)
&=&
\left[
\begin{array}{cc}
{1\over4}-({1\over2}-x)\theta+\theta^2 & \theta^2\sqrt{x(1-x)} \\
\theta^2\sqrt{x(1-x)} & {1\over4}+({1\over2}-x)\theta+\theta^2
\end{array}
\right], 
\\
\hat\phi_2(\theta)
&=& \frac{1-\theta^2}{4}\hat I_\phi.   
\end{eqnarray}
The elements of the risk operator are 
\begin{eqnarray}
\hat W^{(0)}&=&{1\over{27}}
   \left[
         \begin{array}{cc}
         4+9x & 7\sqrt{x(1-x)} \\
         7\sqrt{x(1-x)} & 13-9x
         \end{array}
   \right]
\nonumber\\
&\oplus& {5\over{27}}\hat I_\phi,
\label{case_b_W0}
\\
\hat W^{(1)}&=&{1\over{27}}
   \left[
         \begin{array}{cc}
         7x & 5\sqrt{x(1-x)} \\
         5\sqrt{x(1-x)} & 7(1-x)
         \end{array}
   \right]
\nonumber\\
&\oplus& {1\over{27}}\hat I_\phi,  
\\
\hat W^{(2)}
&=&{1\over{405}}
   \left[
         \begin{array}{cc}
         4+75x & 61\sqrt{x(1-x)} \\
         61\sqrt{x(1-x)} & 79-75x
         \end{array}
   \right]
\nonumber\\
&\oplus&\frac{11}{405}\hat I_\phi.   
\end{eqnarray}
The minimizing operator is (see Appendix \ref{ap_b}) 
\be\label{Theta_b}
\hat\Theta
=\frac{1}{17[13+8x(1-x)]}
   \left[
         \begin{array}{cc}
         a & c \\
         c & b
         \end{array}
   \right]
   \oplus {1\over5}\hat I_\phi,  
\ee
where 
\be
\begin{array}{lll}
a&=&7x(35-20x+2x^2), \\
b&=&7x(17+16x+2x^2), \\
c&=&9[9-2x(1-x)]\sqrt{x(1-x)}.  
\end{array} 
\ee
To diagonalize it we use 
$r=\sqrt{(a-b)^2+4c^2}$ and 
\be
\mathrm{cos}\gamma=\sqrt{\frac{r+a-b}{2r}}, 
\quad
\mathrm{sin}\gamma=\sqrt{\frac{r-a+b}{2r}}. 
\ee
We then have the similar eigenstates to Eq. (\ref{eigenvec1}) and 
the eigenvalues $\theta_1=\theta_+$, $\theta_2=\theta_-$, and 
$\theta_3=\theta_4={1\over5}$ with 
\be
\theta_\pm=\frac{119[1+2x(1-x)] \pm r}{34[13+8x(1-x)]}. 
\ee
The average cost is then   
\be
\bar C_2(x)=\frac{8[391+606x(1-x)-10x^2(1-x)^2]}{2295[13+8x(1-x)]}.  
\ee
This is an upward convex function, symmetric with respect to 
$x={1\over2}$. The minimum is attained at $x=0,1$, that is, by  
separable input states. This reads 
$\bar C_{2\mathrm{min}}=\frac{184}{1755}$.  
\begin{figure}
\begin{center}
\includegraphics[width=0.41\textwidth]{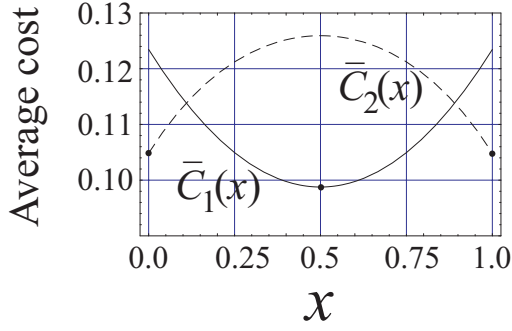}
\end{center}
\caption{\label{fig:Cx}
The average costs as a function of $x$. 
}
\end{figure}
\begin{figure}
\begin{center}
\includegraphics[width=0.43\textwidth]{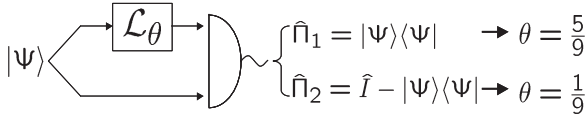}
\end{center}
\caption{\label{fig:scheme}
The optimal estimation strategy using two probe qubits. 
$\ket\Psi$ is the maximally entangled state. The output state is 
projected onto $\{\hat\Pi_1, \hat\Pi_2\}$. We guess the channel 
parameter as $\theta={5\over9}$ for the outcome $\hat\Pi_1$ and 
$\theta={1\over9}$ otherwise. 
}
\end{figure}

The average costs for cases (a) and (b) are shown in 
Fig.~\ref{fig:Cx}: 
$\bar C_1(x)$ (solid line) and $\bar C_2(x)$ (dashed line). 
We see that $\bar C_{1\mathrm{min}}<\bar C_{2\mathrm{min}}$ so that 
the optimal estimation strategy, using two probe qubits, is to 
prepare them as a maximally entangled pair and to input one qubit 
of the pair into the channel keeping the other untouched. 
The estimation is then obtained by applying the two element POM, 
Eq. (\ref{optimal POM}), as described in Case (a). 
This strategy is represented schematically in 
Fig. \ref{fig:scheme}.

When $M$ maximally entangled pairs $\ket\Psi^{\otimes M}$ are 
available, it is best to use them so as to have the output 
$[({\cal L}_\theta\otimes\hat I)\proj{\Psi}]^{\otimes M}$. 
The optimal measurement for this can be derived straightforwardly. 
This is discussed in the next section as a part of an arbitrary 
finite dimensional case.

\section{$d$-dimensional case}
\label{sec:d-dim}

The action of the depolarizing channel on a $d$ dimensional system 
is described by 
\be
{\cal L}_\theta \hat\rho = \theta \hat\rho 
                         + \frac{1-\theta}{d}\hat I. 
\ee
Complete positivity then implies $-{1\over{d^2-1}}\le\theta\le1$. 
For $d\ge3$, we have not succeeded in finding the optimal probe state, 
even when we restrict ourselves to a pure state. 
In this section we focus on the most plausible input state, 
that is, the maximally entangled state, and consider the estimation 
using $M$ entangled pairs. 
Only for $d=2$, is the optimality ensured.

It might be interesting to compare the three cases specified by the 
three different outputs;   
\begin{itemize}
\item[(a)] \enskip
$M$ product states of the pair 
\begin{eqnarray}
\hat\Psi_1(\theta)
    &=&({\cal L}_\theta\otimes\hat I) \proj{\Psi}, 
\nonumber\\
    &=&\theta\proj{\Psi}+\frac{1-\theta}{d^2}\hat I\otimes\hat I,
\label{output-a-1}
\end{eqnarray}
where $\ket{\Psi}$ is the maximally entangled state, 
\item[(b)] \enskip
$M$ product states of the pair 
\begin{eqnarray}
\hat\Psi_2(\theta)
    &=&({\cal L}_\theta\otimes{\cal L}_\theta) \proj{\Psi}, 
\nonumber\\
    &=&\theta^2\proj{\Psi}
      +\frac{1-\theta^2}{d^2}\hat I\otimes\hat I, 
\end{eqnarray}
\item[(c)] \enskip
$2M$ product states of 
\begin{eqnarray}
\hat\psi(\theta)
    &=&{\cal L}_\theta\proj{0}, 
\nonumber\\  
    &=&\theta\proj{0}+\frac{1-\theta}{d}\hat I. 
\end{eqnarray}
\end{itemize}
(The input state in case (c) can be any pure state in the $d$ 
dimensional space.) 
Let us first consider the case (a). 
We denote Eq. (\ref{output-a-1}) as 
\be
\hat\Psi(\theta)=f_0(\theta) \hat a_0 + f_1(\theta) \hat a_1, 
\label{output-a-2}
\ee
where 
\be
\hat a_0\equiv\proj{\Psi}, \quad 
\hat a_1\equiv\hat I-\proj{\Psi}, 
\ee
and 
\be
f_0(\theta)=\theta+\frac{1-\theta}{d^2}, \quad 
f_1(\theta)=\frac{1-\theta}{d^2}. 
\ee
The output state can then be represented as 
\be
\hat\Psi(\theta)^{\otimes M}=\sum_{m=0}^M
f_0(\theta)^{M-m} f_1(\theta)^{m} \hat A_m, 
\ee
where 
\be
\hat A_m=\sum_{(i_1+...i_M=m)}
\hat a_{i_1}\otimes\cdots\otimes\hat a_{i_M}, 
\label{projector-A}
\ee
is the projector onto the symmetric subspace. 
The risk operator is 
\be
\hat W(\theta)=\sum_{m=0}^M
[\omega_m^{(2)}-2\theta\omega_m^{(1)}+\theta^2\omega_m^{(0)}] 
\hat A_m,
\ee
where 
$\omega_m^{(k)}\equiv
\int_{-{1\over3}}^1d\theta 
\theta^k f_0(\theta)^{M-m} f_1(\theta)^{m}$. 
The optimal POM is 
\be
\hat\Pi(\theta)=\sum_{m=0}^M \delta(\theta-\theta_m) \hat A_m,  
\label{POM_M}
\ee 
where 
\be
\theta_m\equiv\frac{\omega_m^{(1)}}{\omega_m^{(0)}}, 
\label{estimate}
\ee
We then note that  
\be
\hat W(\theta)-\hat\Gamma
=\sum_{m=0}^M (\theta-\theta_m)^2 \omega_m^{(0)} \hat A_m\ge0, 
\ee 
from which it can easily be seen that the conditions (i) and (ii) 
hold. 
The minimum average cost is 
\be
\bar C_1(M)=\sum_{m=0}^M
\left[
\omega_m^{(2)}-\frac{(\omega_m^{(1)})^2}{\omega_m^{(0)}}
\right]
\left(\begin{array}{c}
        M\\m
      \end{array}
\right)
(d^2-1)^m. 
\label{C1_M}
\ee

The other cases can be dealt with in a similar manner. 
In the case (b), we just put 
\be
f_0(\theta)=\theta^2+\frac{1-\theta^2}{d^2}, \quad 
f_1(\theta)=\frac{1-\theta^2}{d^2}.
\label{f0f1_C2} 
\ee
The minimum average cost $\bar C_2(M)$ is then given by the same 
expression as Eq. (\ref{C1_M}) with $\omega_m^{(k)}$'s defined by 
$f_0(\theta)$ and $f_1(\theta)$ of Eq. (\ref{f0f1_C2}).

In the case (c), we use 
\be
\hat a_0\equiv\proj{0}, \quad 
\hat a_1\equiv\hat I -\proj{0}, 
\ee
and 
\be
f_0(\theta)=\theta+\frac{1-\theta}{d}, \quad 
f_1(\theta)=\frac{1-\theta}{d}.
\label{f0f1_Cs} 
\ee
The minimum average cost is 
\be
\bar C_\mathrm{SEP}(M)=\sum_{m=0}^{2M}
\left[
\omega_m^{(2)}-\frac{(\omega_m^{(1)})^2}{\omega_m^{(0)}}
\right]
\left(\begin{array}{c}
        2M\\m
      \end{array}
\right)
(d-1)^m. 
\label{Csepmin_M}
\ee

\begin{figure}
\begin{center}
\includegraphics[width=0.41\textwidth]{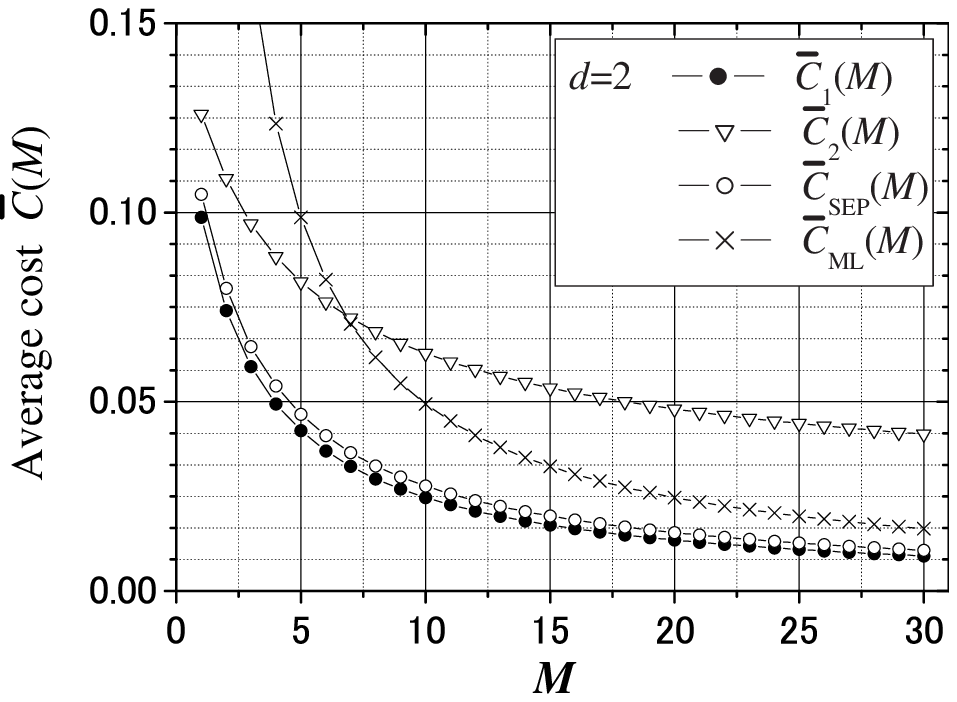}
\end{center}
\caption{\label{fig:CMd2}
The average costs as a function of the number of pairs. 
}
\end{figure}

\begin{figure}
\begin{center}
\includegraphics[width=0.41\textwidth]{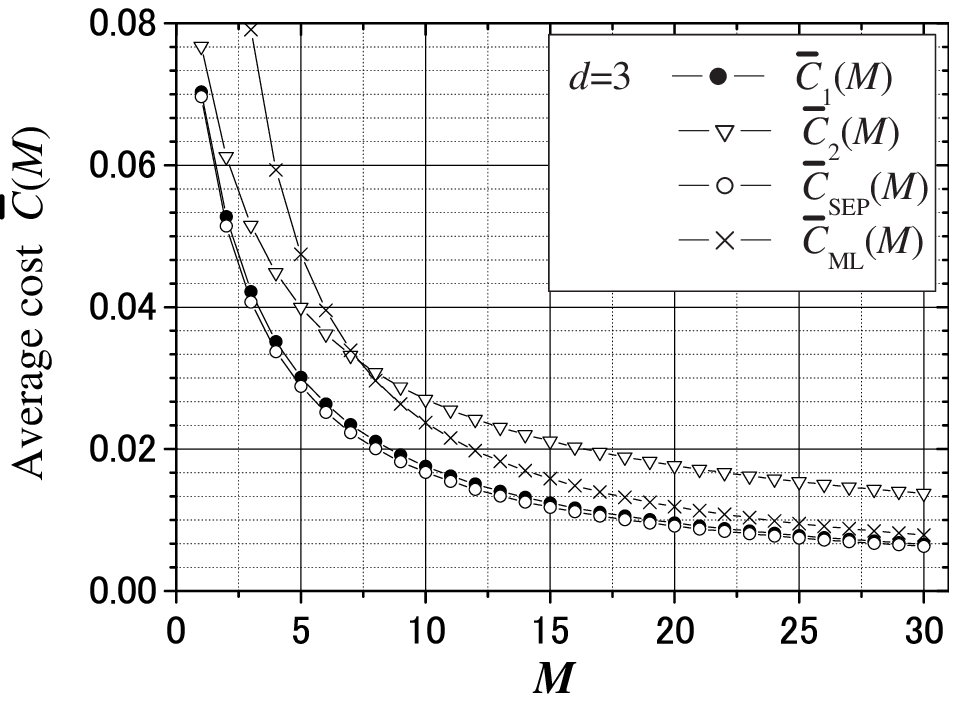}
\end{center}
\caption{\label{fig:CMd3}
The average costs as a function of the number of pairs. 
}
\end{figure}

\begin{figure}
\begin{center}
\includegraphics[width=0.41\textwidth]{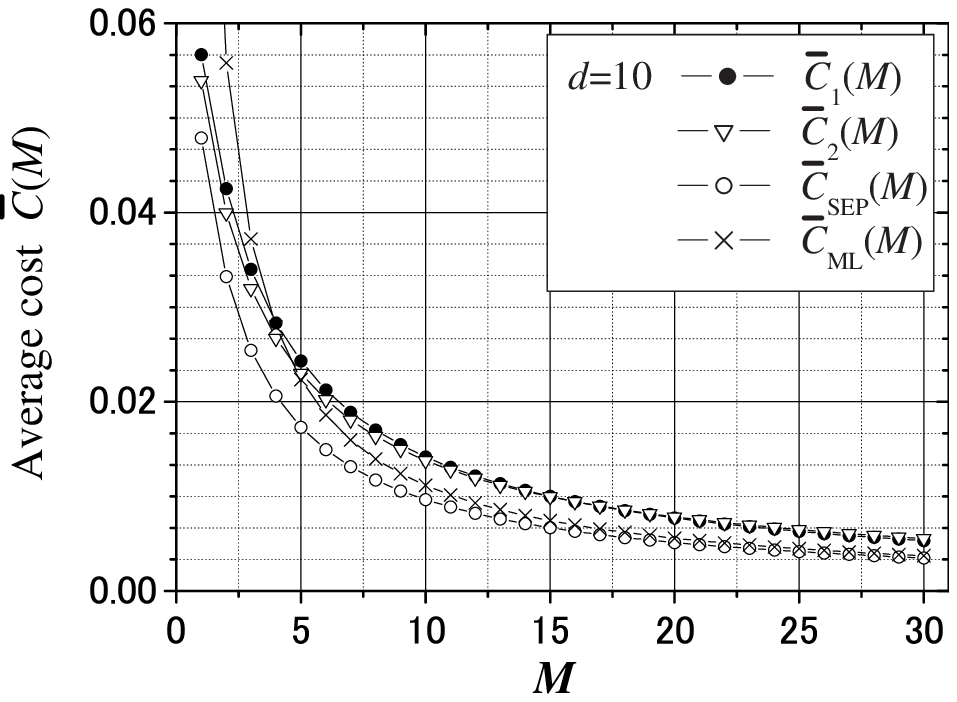}
\end{center}
\caption{\label{fig:CMd10}
The average costs as a function of the number of pairs. 
}
\end{figure}

The three costs $\bar C_1(M)$, $\bar C_2(M)$, and 
$\bar C_\mathrm{SEP}(M)$ 
are plotted in Fig. \ref{fig:CMd2} $(d=2)$, 
Fig. \ref{fig:CMd3} $(d=3)$, and Fig. \ref{fig:CMd10} $(d=10)$. 
In the figures another average cost $\bar C_\mathrm{ML}(M)$ 
is also plotted. 
This cost is by the strategy belonging to the case (c), but unlike 
the one attaining $\bar C_\mathrm{SEP}(M)$, the estimator is made by 
the maximum likelyhood principle for which 
\be
\theta_m=\frac{md}{2M(d-1)}, 
\ee
instead of Eq. (\ref{estimate}), and leads to the analytic expression 
\be
\bar C_\mathrm{ML}(M)=\frac{1}{2M}\frac{d^5(d+3)}{6(d^2-1)^3}
\ee
It is this strategy that was used in ref. \cite{Cirone01} for the 
case of $d=2$.

For $d\ge3$, the minimum average cost is always attained by a 
separable probe state. 
Only in the two dimensional case, is it the bipartite entangled probe 
that attains the minimum average cost. 
It is worth mentioning the depolarizing channel with the narrower 
parameter region $0\le\theta\le1$, which is a more commonly 
used model with an well defined interpretation of randomized 
\textit{probability} of $\theta$. 
We found that the best probe in this model is always a separable 
state. 
In this sense a separable state is generally an adequate probe state 
for the depolarizing channel estimation as far as the comparison with 
a bipartite entangled probe state is concerned.


\section{Concluding remark}
\label{sec:remark}

When we have several identical samples at our disposal, 
it might be desirable to apply the best 
\textit{collective} measurement on the whole system. 
This means preparing a single multi-qubit state followed by an 
optimized measurement. 
We might also consider performing 
a preliminary measurement on a part of the system and then 
feedback this back to deal with the remaining part. 
But in the case of the previous section, the collective measurement on 
$M$ identical output pairs or $2M$ identical output particles is not 
necessary. 
The action of the depolarizing channel on a maximally entangled 
state always results in a statistical mixture between 
the input state and its orthogonal complement  
(Eq. (\ref{output-a-2})). 
Estimating the channel parameter is nothing but determining this 
mixing ratio, which is a \textit{classical} distribution. 
Therefore the optimal measurement is realized by a separable type 
constructed 
by the binary orthogonal projectors $\{\hat a_0, \hat a_1 \}$ 
according to Eq. (\ref{projector-A}). 
In the case where the output state includes the channel parameter as 
a quantum distribution, that is, the parameter appears in the off 
diagonal components in the density matrix, the optimal measurement 
would be a collective measurement. 
When the channel includes a unitary opreration, we will have to face 
this problem. Channel estimation for such a case is a future problem.

It is a remaining problem to see how effective the multipartite 
entangled probe is. 
However, in the estimation of decoherence channel under the power 
constraint scenario, 
that is, under a given and fixed number of probe particles, 
it seems more common that entanglement is not necessary. 
In fact, in the cases of the amplitude damping channel and 
dephasing channel, there is no merit to use entangled probe. 
In the amplitude damping channel, for example, 
the best probe is to input the most highly excited state. 
An entangled probe is rather wasteful because this includes the state 
components other than the excited state and these components are less 
sensitive to the damping.

Finally it might be interesting to study the multi parameter case, 
such as the Pauli channel estimation. 
We may then ask how to optimaize (in Bayesian sense) the simultaneous 
measurement on the noncommuting observables as well as searching for 
appropriate probe states.

%
\begin{acknowledgments}
We are grateful to Mr. K. Usami, Dr. Y. Tsuda, and Dr. K. Matsumoto 
for helpful discussions. 
This work was supported, in part, by the British Council,
the Royal Society of Edinburgh, and by the Scottish
Executive Education and Lifelong Learning Department.
\end{acknowledgments}
%
\appendix*
\section{Derivation of Eq. (\ref{Theta_a})}
\label{app_a}

For obtaining the minimizing operator $\Theta$ in Eq. (\ref{Theta_a}), 
we first diagonalize $\hat W^{(0)}$ by 
$\hat U_0=\hat u_0\oplus\hat I_\phi$ where 
\be
\hat u_0
=  \left[
         \begin{array}{cc}
         \mathrm{cos}\gamma_0 & -\mathrm{sin}\gamma_0 \\
         \mathrm{sin}\gamma_0 & \mathrm{cos}\gamma_0
         \end{array}
   \right],  
\ee
with $r_0=\sqrt{1-3x(1-x)}$ and 
\be
\mathrm{cos}\gamma_0=\sqrt{\frac{r_0-1+2x}{2r_0}}, 
\quad
\mathrm{sin}\gamma_0=\sqrt{\frac{r_0+1-2x}{2r_0}}. 
\ee
The spectral decomposition 
\be
\hat W^{(0)}=\sum_{i=1}^4 \omega_i \proj{\omega_i}. 
\ee
is given by
\be\label{eigenvec0}
\begin{array}{lll}
\ket{\omega_1}&=\hat u_0\ket{\mu_1}, \quad
              &\omega_1=(1+r_0)/3, \\
\ket{\omega_2}&=\hat u_0\ket{\mu_2}, \quad
              &\omega_2=(1-r_0)/3, \\
\ket{\omega_3}&=\ket{\nu_1}, \quad
              &\omega_3=(1-x)/3, \\
\ket{\omega_4}&=\ket{\nu_2}, \quad
              &\omega_4=x/3. 
\end{array} 
\ee
We then calculate 
\be
\tilde\Theta=\sum_{i,j=1}^4 \frac{2}{\omega_i+\omega_j} 
\ket{i}\bra{\omega_i} \hat W^{(1)} \ket{\omega_j}\bra{j},  
\label{minmizing_op}
\ee
where 
\be
\begin{array}{lll}
\ket{1}&=\ket{\mu_1}, \\
\ket{2}&=\ket{\mu_2}, \\
\ket{3}&=\ket{\nu_1}, \\
\ket{4}&=\ket{\nu_2}. 
\end{array} 
\ee
This gives  
\be
\tilde\Theta=
  \tilde\Theta_\psi \oplus {1\over9}\hat I_\phi,  
\ee
where 
\be
\tilde\Theta_\psi
= {1\over9} 
\left[
\begin{array}{cc}
\frac{4r_0(1+r_0)+3x(1-x)}{r_0(1+r_0)} & 
     -\frac{3(1-2x)\sqrt{x(1-x)}}{r_0} \\
-\frac{3(1-2x)\sqrt{x(1-x)}}{r_0} & 
     \frac{4r_0(1-r_0)-3x(1-x)}{r_0(1-r_0)}
\end{array}
\right]. 
\ee
The minimizing operator is given by 
$\hat\Theta=\hat U_0\tilde\Theta\hat U_0^\dagger$ 
which results in Eq. (\ref{Theta_a}).

\section{Derivation of Eq. (\ref{Theta_b})}
\label{app_b}

The unitary operator for diagonalizing $\hat W^{(0)}$ in 
Eq. (\ref{case_b_W0}) is 
$\hat U_0=\hat u_0\oplus\hat I_\phi$ where 
\be
\hat u_0
=  \left[
         \begin{array}{cc}
         \mathrm{cos}\gamma_0 & -\mathrm{sin}\gamma_0 \\
         \mathrm{sin}\gamma_0 & \mathrm{cos}\gamma_0
         \end{array}
   \right],  
\ee
with $r_0=\sqrt{81-128x(1-x)}$ and 
\be
\mathrm{cos}\gamma_0=\sqrt{\frac{r_0-9(1-2x)}{2r_0}}, 
\quad
\mathrm{sin}\gamma_0=\sqrt{\frac{r_0+9(1-2x)}{2r_0}}. 
\ee
The spectral decomposition 
\be
\hat W^{(0)}=\sum_{i=1}^4 \omega_i \proj{\omega_i}. 
\ee
is given by
\be\label{eigenvec0}
\begin{array}{lll}
\ket{\omega_1}&=\hat u_0\ket{\mu_1}, \quad
              &\omega_1=(17+r_0)/54, \\
\ket{\omega_2}&=\hat u_0\ket{\mu_2}, \quad
              &\omega_2=(17-r_0)/54, \\
\ket{\omega_3}&=\ket{\nu_1}, \quad
              &\omega_3=5/27, \\
\ket{\omega_4}&=\ket{\nu_2}, \quad
              &\omega_4=5/27. 
\end{array} 
\ee
We then have 
\be
\tilde\Theta=\tilde\Theta_\psi \oplus {1\over5}\hat I_\phi,  
\ee
where 
\be
\tilde\Theta_\psi
=  
\left[
\begin{array}{cc}
\frac{7[r_0+9-16x(1-x)]}{r_0(17+r_0)} & 
     \frac{8(1-2x)\sqrt{x(1-x)}}{17r_0} \\
\frac{8(1-2x)\sqrt{x(1-x)}}{17r_0} & 
     \frac{7[r_0-9+16x(1-x)]}{r_0(17-r_0)}
\end{array}
\right]. 
\ee
Substituting this to 
$\hat\Theta=\hat U_0\tilde\Theta\hat U_0^\dagger$, we have  
Eq. (\ref{Theta_b}).

%
%

%
%
\end{document}